\documentclass[%
 aip,
 amsmath,amssymb,
 reprint,%
]{revtex4-1}

\usepackage{graphicx}
\usepackage{dcolumn}
\usepackage{bm}

\usepackage[utf8]{inputenc}
\usepackage[T1]{fontenc}
\usepackage{mathptmx}
\usepackage{etoolbox}
\usepackage{xcolor}

\makeatletter
\def\@email#1#2{%
 \endgroup
 \patchcmd{\titleblock@produce}
  {\frontmatter@RRAPformat}
  {\frontmatter@RRAPformat{\produce@RRAP{*#1\href{mailto:#2}{#2}}}\frontmatter@RRAPformat}
  {}{}
}%
\makeatother
\begin{document}

\preprint{AIP/123-QED}

\title[Phase Control of Nonlinear Breit-Wheeler Pair Creation]{Phase Control of Nonlinear Breit-Wheeler Pair Creation}
\author{B. Barbosa}
\email{bernardo.barbosa@tecnico.ulisboa.pt}
 \affiliation{ GoLP/Instituto de Plasmas e Fusão Nuclear, Universidade de Lisboa, Lisbon, Portugal 
}%

\author{J. P. Palastro}
 \affiliation{%
Laboratory for Laser Energetics, University of Rochester, Rochester, NY, USA
}

\author{D. Ramsey}
\affiliation{%
Laboratory for Laser Energetics, University of Rochester, Rochester, NY, USA
}

\author{K. Weichman}
\affiliation{%
Laboratory for Laser Energetics, University of Rochester, Rochester, NY, USA
}%

\author{M. Vranic}%
\affiliation{ GoLP/Instituto de Plasmas e Fusão Nuclear, Universidade de Lisboa, Lisbon, Portugal 
}%

\date{\today}

\begin{abstract}
Electron-positron pair creation occurs throughout the universe in the environments of extreme astrophysical objects, such as pulsar magnetospheres and black hole accretion disks. The difficulty of emulating these environments in the laboratory has motivated the use of ultrahigh-intensity laser pulses for pair creation. Here we show that the phase offset between a laser pulse and its second harmonic can be used to control the relative transverse motion of electrons and positrons created in the nonlinear Breit-Wheeler process. Analytic theory and particle-in-cell simulations of a head-on collision between a two-color laser pulse and electron beam predict that with an appropriate phase offset, the electrons will drift in one direction and the positrons in the other. The resulting current may provide a collective signature of nonlinear Breit-Wheeler, while the spatial separation resulting from the relative motion may facilitate isolation of positrons for subsequent applications or detection.

\end{abstract}


\maketitle

\section{Introduction}
\label{sec:intro}

The environment of extreme astrophysical objects feature electromagnetic fields with sufficient strength to convert photons into electron-positron pairs.\cite{Peter1969,Goldreich1969,Sturrock1971,Ruderman1975,Damour1975,Gibbons1975,Arons1979,Takahara1985,Henri1991,Beloborodov1999,Ruffini2010} The large-scale production of pairs forms a dense pair plasma that plays a critical role in the dynamics of the environment. While observable signals from the pair plasma, such as $\gamma$-ray bursts,\cite{Uso1992,Meszaros2000,vanPutten2000,Peer_2004,Siegert2016} provide valuable insights into these dynamics, the complex interplay of physical processes within the environment make it difficult to study any single phenomenon in isolation. By emulating the conditions of extreme astrophysical objects, laboratory experiments can isolate phenomena and validate the simulation tools used for their study. Relativistic particle beams and ultrahigh-intensity laser-matter interactions offer two approaches to these experiments. \cite{Rafelski1978,DiPiazza2012,Gonoskov2022,Brandenburg2023}

The strong-field quantum electrodynamical (QED) processes responsible for pair creation are at the current frontier of high-intensity laser-matter interactions.\cite{SLAC,Bulanov2010,Chen,Sarri,Vranic2018,Blackburn2018,Mercuri-Baron2021,Matheron2023} Creative configurations for these interactions allow for pair production at field strengths well below the characteristic electric (magnetic) field of nonlinear QED, i.e., the Sauter-Schwinger field equal to $10^{18}~\mathrm{V/m}$ ($10^9~\mathrm{T}$).\cite{Schwinger,Furry1951} For instance, irradiating a thin, high-atomic number target with an intense laser pulse can energize a population of electrons. The bremsstrahlung photons emitted by the electrons subsequently decay into pairs through the Bethe-Heitler process.\cite{Bethe-Heitler,Sarri,Chen} As another example, the combined fields of a plasma and laser pulse channeling through a thin, dense target can rapidly accelerate electrons, resulting in both forward and backward photon emission. The collision of the counter-propagating photons can produce pairs through the linear Breit-Wheeler process. \cite{Breit-Wheeler,Sugimoto2023}

The nonlinear Breit-Wheeler process can be used to generate pairs without a dense target in near-vacuum conditions. In a typical configuration, an ultrahigh-intensity laser pulse collides head on with relativistic electrons.\cite{SLAC} Nonlinear Compton scattering (NCS) of low-energy optical photons from the electrons produces high-energy photons. Immersion of the high-energy photons in the fields of the intense laser pulse allows for their decay into electron-positron pairs, i.e., Breit-Wheeler pair creation. Under the right conditions, subsequent NCS from the pairs can ignite a QED cascade, where the number of leptons grows exponentially---The ultimate result of which is a pair plasma. \cite{Bulanov2006,Bell2008,Kirk2009,Bulanov2010,Bulanov2010-2,Nerush2011,Zhidkov2014,Grismayer2016,Vranic2016,Mironov2016,Jirka2016,Kostyukov2016,Grismayer2017,Jirka2017}

Regardless of the scheme used for pair creation, several effects can challenge reliable diagnosis and analysis of experimental results. These include direct trident pair creation\cite{Trident} or bremsstrahlung and Bethe-Heitler byproducts from the unintended interaction of the laser pulse or charged particles with experimental equipment. Methods for separating the electrons from the positrons in a predictable way can facilitate detection and diagnosis of the pairs. Conventionally, this is done by deflecting both species in a static magnetic field (on the order of 1 T).\cite{SLAC,Sarri,Chen} However, this technique separates the pairs far from the interaction region. As an alternative, the field of an ultrahigh-intensity laser pulse can be structured to deflect the electrons and positrons in opposite directions at the instant of their creation.

\begin{figure*}[!ht]
\centering
\includegraphics[width=0.7\textwidth]{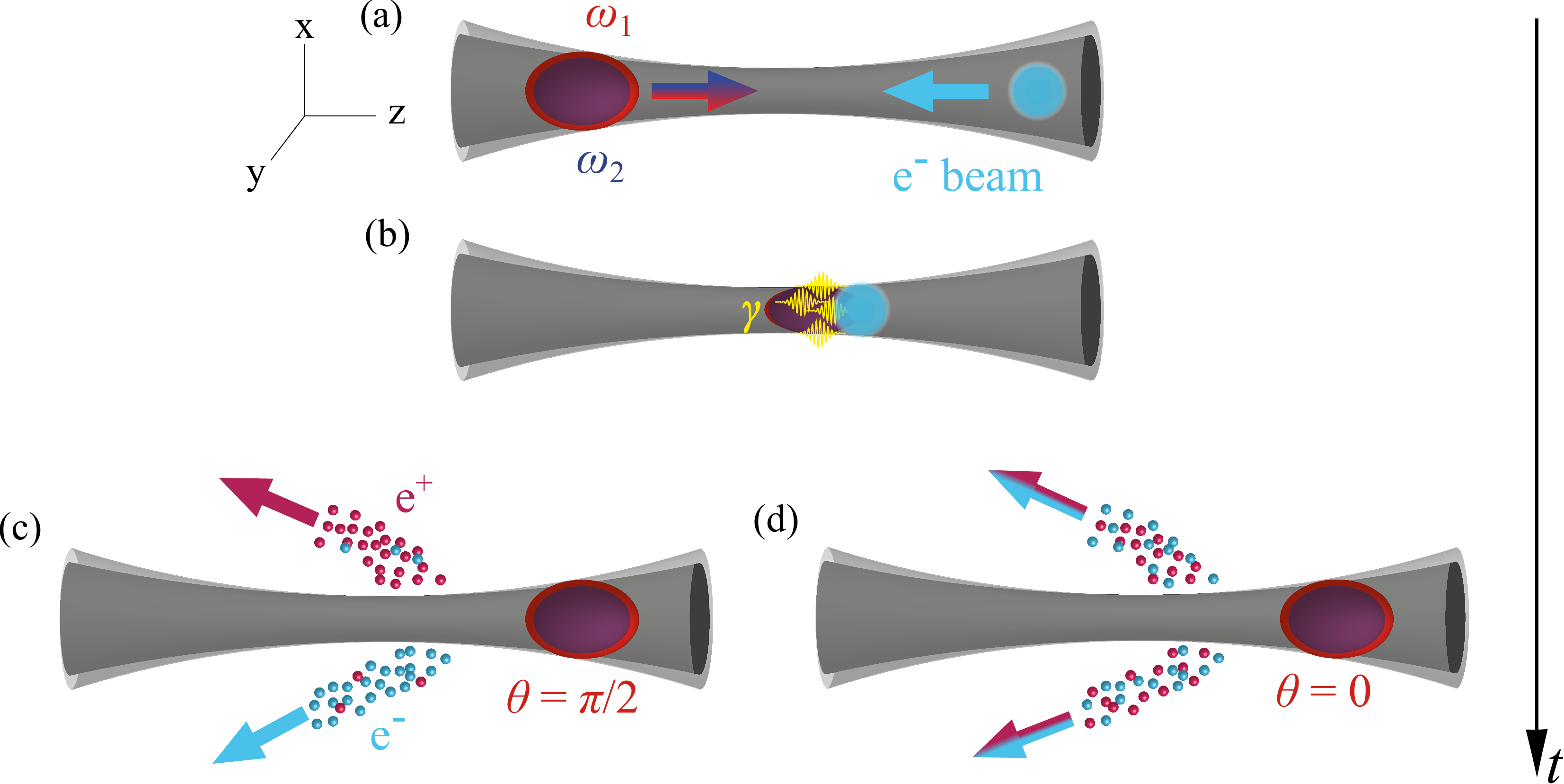}
\caption{\label{fig:setup} Schematic of the electron beam-laser pulse interaction for controlling the relative motion of electrons and positrons created by the nonlinear Breit-Wheeler process. (a) The electron beam and two-color laser pulse right before their head-on collision. The harmonics composing the laser pulse have the same focused spot size and focal plane. The transverse width of each harmonic along its propagation path is displayed in gray.  (b) Nonlinear Compton scattering during the collision of the electrons and laser pulse produces hard photons ($\gamma$-rays) shown in yellow. The interaction of the hard photons with the laser pulse produces electron-positron pairs. (c) For a phase offset between the harmonics $\theta=\pi/2$, the electrons and positrons drift in opposite directions. (d) For a phase offset $\theta=0$, the electrons and positrons drift in both directions.}
\end{figure*}

Here we demonstrate that the relative transverse motion of electrons and positrons created in the nonlinear Breit-Wheeler process can be controlled using a laser pulse composed of a fundamental and an appropriately phased second harmonic. By adjusting the relative phase of the harmonics, the electrons can be made to drift in one direction and the positrons in the other (Fig. \ref{fig:setup}). The presence of the second harmonic (or any even harmonic) breaks the symmetry of the field and allows the charges to drift in opposite directions. This cannot be achieved using odd harmonics alone. The two-color field configuration is motivated by techniques for coherent control used in photoionization, photoemission, molecular orientation, radiation reaction, and electron-positron spin polarization. \cite{Vrakking1997,Two-color,Control,Forster2016,Maroju2023,DiPiazza2014,Spin} 

The remainder of this manuscript is organized as follows. Section II describes the pair-creation scheme and presents an analytical model for the pair motion and current. Comparisons with 1D QED-particle-in-cell (QED-PIC) simulations\cite{Dawson,OSIRIS,VranicThesis,ClassicalRR} of a two-color laser pulse colliding with hard photons show that the model successfully predicts the scaling of the pair current with the relative phase and amplitude of the harmonics. Section III presents full-scale, 2D QED-PIC simulations of a focused laser pulse colliding with a relativistic electron beam for parameters relevant to near-term, high-power laser facilities. Despite the ponderomotive force of the focused pulse, the relative phase of the harmonics still provides control over the relative transverse motion of the created electrons and positrons. Section IV summarizes the results and discusses future prospects. 

\section{Pair Creation in Two-color Laser Pulses}
\label{sec:backg}

Figure~\ref{fig:setup} illustrates the configuration for pair creation. A relativistic electron beam (blue) collides head on with a laser pulse composed of a first and second harmonic (red and dark blue ellipsoids) [Fig.~\ref{fig:setup}(a)]. Rapid acceleration of the electrons in the fields of the laser pulse results in the emission of hard photons ($\gamma$-rays) through nonlinear Compton scattering [Fig.~\ref{fig:setup}(b)]. These photons counter-propagate with respect to the laser pulse, simultaneously interact with multiple optical photons, and decay into pairs through the nonlinear Breit-Wheeler process [Fig.~\ref{fig:setup}(c) and (d)]. 

The pair creation rate and subsequent particle dynamics depend on the phase of the optical field at the instant of creation and the phase offset $\theta$ between the two harmonics. The phase at the instant of creation determines the local field strength, while the phase offset determines the shape of the optical waveform. Tuning the phase offset to structure the optical waveform provides control over the relative transverse motion of the created electrons and positrons [Fig.~\ref{fig:setup}(c) and (d)].

The relative motion of the electrons and positrons can be described by an analytical model for the evolution of the pairs in the electromagnetic fields of the laser pulse. The laser pulse propagates in the positive $\bold{\hat{z}}$ direction and is polarized in the $\bold{\hat{x}}$ direction. The normalized vector potential $\text{\textbf{a}}=e\text{\textbf{A}}/m_ec$ of the pulse is modeled as a plane wave:
\begin{equation}
\text{\textbf{a}}=\text{a}_1(\phi)\sin{\phi}\bold{\hat{x}}+\text{a}_2(\phi)\sin{(2\phi+\theta)}\bold{\hat{x}},
\label{eq:laser_field}
\end{equation}
where $\phi=t-z$ is the phase variable and $\text{a}_1$ and $\text{a}_2$ describe the temporal profiles of the first and second harmonic with maximum values $a_1$ and $a_2$. Here and throughout, time and space are normalized to $1/\omega_1$ and $c/\omega_1$, where $\omega_1$ is the angular frequency of the first harmonic. 

After creation, the electrons and positrons evolve according to the  relativistic equations of motion\cite{jackson}
\begin{equation}\begin{split}
\frac{d\text{\textbf{p}}}{dt} &= \mp\partial_t \text{\textbf{a}}\pm\frac{\text{\textbf{p}}}{\gamma}\times(\boldsymbol{\nabla}\times\text{\textbf{a}})~,\\
\frac{d\gamma}{dt} &= \mp\partial_t \text{\textbf{a}}\cdot\frac{\text{\textbf{p}}}{\gamma}~,
\end{split}
\label{eq:motion}
\end{equation}
where $\text{\textbf{p}}=\gamma\boldsymbol{\beta}$ is the momentum normalized to $m_ec$, $\boldsymbol{\beta}$ the velocity normalized to $c$, and $\gamma=(1+p^2)^{1/2}$ the Lorentz factor. The upper and lower signs are taken for positrons and electrons, respectively. For a vector potential that depends solely on the coordinates in the combination $\phi = t-z$ [as in Eq. \eqref{eq:laser_field}], the equations of motion admit the conservation laws
\begin{equation}\begin{split}
&\frac{d}{dt}\left(p_x\pm \text{a}_x\right) = 0~, \\
&\frac{d}{dt}\left(p_y\right) = 0, \\
&\frac{d}{dt}(\gamma-p_z) = 0~,
\end{split}
\label{eq:conservations}
\end{equation}
where the subscripts $x$, $y$, and $z$ indicate the spatial components of the vector quantities. 

\begin{figure}[!ht]
\centering
\includegraphics[width=\columnwidth]{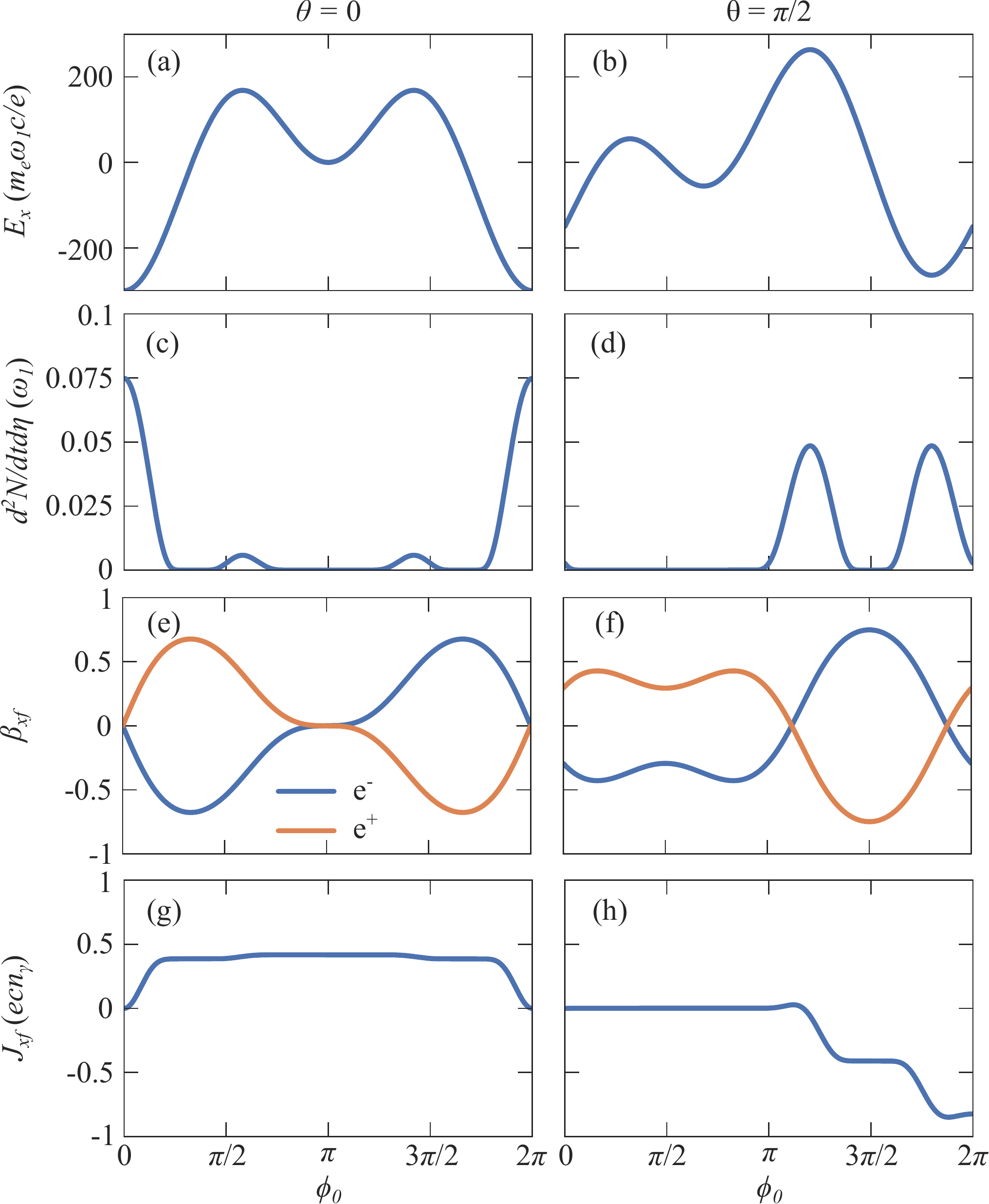}
\caption{\label{fig:drifts} (a,b) The electric field of the two-color laser pulse, (c,d) differential pair creation rate, (e,f) asymptotic drift velocity $\beta_{xf}$ (for created positrons in orange and electrons in blue), and (g,h) cumulative, asymptotic transverse current density of the created pairs as a function of $\phi_0$ for a relative phase of $\theta =0$ (left) and $\theta = \pi/2$ (right). The laser pulse is comprised of a fundamental and second harmonic with wavelengths $\lambda_1 = 2\pi c/\omega_1 = 1~\mathrm{\mu m}$ and $\lambda_2 = 0.5~\mathrm{\mu m}$ and amplitudes $a_1=150$ and $a_2=75$. The incident hard photon had $\hbar\omega/m_ec^2=500$, and its energy was divided evenly between the created electron and positron. When $\theta = \pi/2$, the asymmetric drift velocity and pair creation rate result in the accumulation of a transverse current.}
\end{figure}

Using the conservation laws, the velocity of each particle as it leaves the vicinity of the laser pulse $\boldsymbol{\beta}_f$ can be expressed in terms of its momentum upon creation $\bold{p}_0$ and the vector potential at the instant of its creation a$_{x0}$. More specifically, the transverse velocity asymptotes to
\begin{equation}
\beta_{xf} = \frac{p_{xf}}{\gamma_f} = \frac{2(\gamma_0-p_{z0})(p_{x0}\pm \text{a}_{x0})}{1+(\gamma_0-p_{z0})^2 + (p_{x0}\pm \text{a}_{x0})^2},
\label{eq:drift}
\end{equation}
and $\beta_{yf} = \beta_{y0}$, where the subscript $0$ indicates the value of a quantity at the time of creation. Equation \eqref{eq:drift} demonstrates that after interacting with the laser pulse, the particles drift in the transverse direction with a velocity determined by their charge and the phases of the harmonics at the time of their creation, $\phi_0$ and $2\phi_0 + \theta$ [Figs. \ref{fig:drifts}(e)--(f)].

The initial momenta of an electron-positron pair are determined by the energy and momentum of the hard photon responsible for their creation. In the configuration of interest here (Fig.~\ref{fig:setup}), the incident electron beam is composed of ultra-relativistic electrons with $-p_z \approx \gamma \gg |\bold{a}|$. As a result, the hard photons produced by NCS travel predominantly in the negative $\bold{\hat{z}}$ direction with $k^{\mu}\approx(\omega,0,0,-\omega)$, where $\omega$ is the angular frequency of the hard photon. When a photon decays, its momentum and energy are divided between the resulting electron ($e$) and positron ($p$), such that $p_{x0} \approx 0$ and $\gamma^e_{0} + \gamma^p_{0} = \omega/a_s$, where $a_s = m_ec^2/\hbar\omega_1$ is the normalized vector potential corresponding to the Schwinger field. \cite{Ridgers,VranicThesis,Gonoskov2015}

The partitioning of the hard photon energy between the electron and positron is determined probabilistically by the differential rate for nonlinear Breit-Wheeler pair creation $d^2N_{BW}/dtd\eta$. \cite{Rates1,Rates2,Rates3} Here, $\eta = |(F_{\mu\nu} p^{\mu})^2|^{1/2}/a_s$ is the Lorentz-invariant quantum nonlinearity parameter, $F_{\mu\nu}$ is the electromagnetic tensor (normalized to $\omega_1m_e/e$), and $p^{\mu}$ is the momentum 4-vector of the created electron. At the instant of creation, 
\begin{equation}
\eta(\gamma_0,\phi_0)=2\gamma_0\frac{|\text{a}_1(\phi_0)\cos{(\phi_0)}+2\text{a}_2(\phi_0)\cos{(2\phi_0+\theta)}|}{a_s}~.
\label{eq:eta_birth}
\end{equation}
The differential rate is a sensitive function of $\eta$, and thus of the phases $\phi_0$ and $\theta$. 

Figures \ref{fig:drifts}(a)--(f) illustrate the dependence of the differential pair creation rate $dN_{BW}/dtd{\eta}$ and the asymptotic drift velocity $\beta_{xf}$ on $\phi_0$ and $\theta$. The maxima of the rate coincide with extrema of the electric field, $\bold{E} = -\partial_t \bold{a}$. When $\theta = 0$, the drift is anti-symmetric and alternates every half-cycle [Fig. \ref{fig:drifts}(e)]. When $\theta = \pi/2$, the drift is \textit{asymmetric}, with a larger magnitude in one half of the cycle than the other [Fig. \ref{fig:drifts}(f)]. The combination of the asymmetric drift and asymmetric pair creation rate [Fig. \ref{fig:drifts}(d)] result in a net electron-positron current [Fig. \ref{fig:drifts}(h)]. 

For a mono-energetic beam of hard photons, the asymptotic transverse current density of created pairs is given by
\begin{equation}
J_{xf}=2e\int\int_{\eta_{min}}^{\eta_{max}}\beta_{xf}\frac{d^2N_{BW}}{dtd\eta}n_{\gamma}d\eta d(\phi_0/2)~,
    \label{eq:current}
\end{equation}
where $n_{\gamma}$ is the number density of photons, $\eta_{\text{min}} = \eta(1,\phi_0)$, and $\eta_{\text{max}} = \eta(\omega/a_s-1,\phi_0)$. The factor of 2 accounts for the equal contribution to the current from positrons and electrons, while the factor of $1/2$ accounts for counter-propagation of the hard photons with respect to the laser pulse. Figures \ref{fig:drifts}(g) and (h) show the cumulative integral of the current density over a single phase of a two-color laser field.

\begin{figure}
\includegraphics[width=\columnwidth]{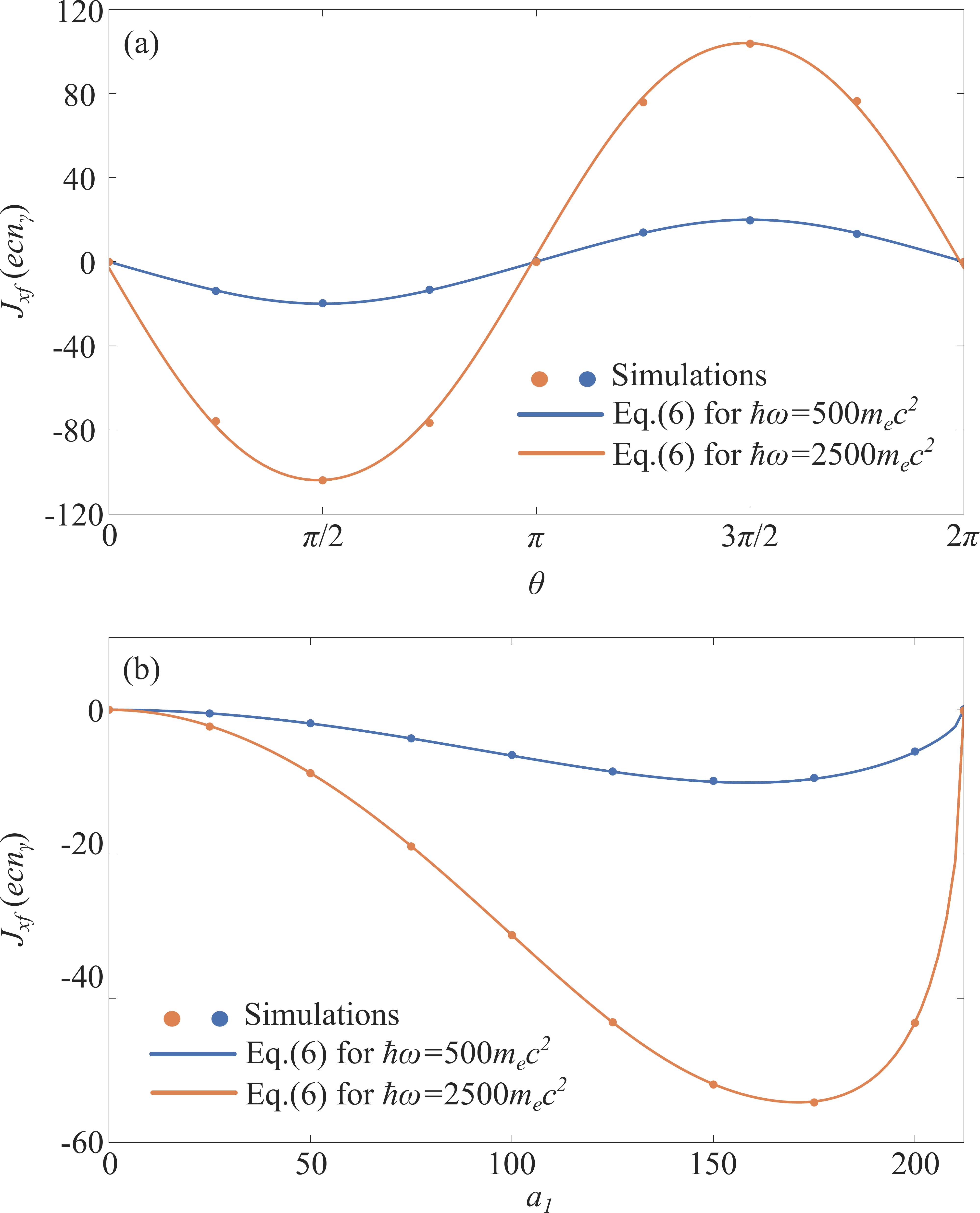} 
    \caption{Comparison of the asymptotic transverse current density predicted by the analytical theory (continuous curves) and 1D simulations (dots) as a function of (a) the phase offset between the first and second harmonic of the laser pulse $\theta$ and (b) the amplitude of the first harmonic $a_1$ for a fixed fluence of $1.4\times10^9 \, \mathrm{J/cm^2}$ and a full-width-half-maximum duration of $23 \,\mathrm{fs}$. In (a) $a_{1}=150$, while in (b) $\theta=\pi/2$. The theory and simulations are in excellent agreement.}
    \label{fig:comparisons}
\end{figure}

A phase offset of $\theta = \pi/2$ produces the largest relative drift between the created electrons and positrons and maximizes the transverse current density [Fig. \ref{fig:comparisons}(a)]. This prediction of the analytical model agrees with 1D QED-PIC simulations of planar laser pulses colliding head on with monoenergetic, hard-photon beams (see Appendix A for details). Both the model and simulations also predict that the current density is maximized for $a_1 \approx 160$ and $a_2 \approx 70$ when the total energy (fluence in 1D) of the laser pulse is held fixed [Fig. \ref{fig:comparisons}(b)]. This maximum is the result of two effects. First, in the limit that either harmonic has all of the energy, the asymmetry in the field is eliminated, and there is no asymptotic transverse current [$a_1 = 0$ and $210$ in Fig. \ref{fig:comparisons}(b)]. Second, the maximum electric field of the total waveform, and thus the maximum creation rate, occurs when the electric field strengths of the harmonics are equal ($a_1 \approx 150$ in Fig. \ref{fig:comparisons}).

The differential pair creation rate and asymptotic current increase nonlinearly with the energy of the hard photons up to a value of $\chi \gtrsim 1$, where $\chi = 2\omega|\bold{E}|/a_s^2$ is the Lorentz-invariant quantum nonlinearity parameter of the photons.
When the photon energy is large enough ($\chi \sim 1$), prolific creation of pairs significantly depletes the number of available photons. More specifically, the photon number density evolves according to
\begin{equation}
n_{\gamma}(\phi_0) = n_{\gamma i}\exp\left[-\int_{-\infty}^{\phi_0}\int_{\eta_{\mathrm{min}}}^{\eta_{\mathrm{max}}}\frac{d^2N_{BW}}{dtd\eta}d\eta d(\phi_0/2) \right],
\label{eq:depletion}
\end{equation}
where $n_{\gamma i}$ is the density of the incident photon beam. For Fig. \ref{fig:comparisons}, this correction to the photon density was applied in Eq. \eqref{eq:current}, resulting in excellent agreement between the theory and simulations.


\section{Phase Control with Focused Laser Pulses}
\label{sec:resul}

The model and simulations presented in the previous section illustrated the salient phenomena that allow for phase control of nonlinear Breit-Wheeler pair creation. Several effects, however, were not considered: the generation of non-monoenergetic photons from NCS, radiation reaction, and the transverse ponderomotive force of the laser pulse. This section presents full-scale 2D QED-PIC simulations that include these effects and verify that the underlying physical picture remains unchanged.

\begin{table}[b]
\caption{\label{tab:table1}%
Parameters for the 2D QED-PIC simulations. Where applicable, the parameters are the same as in the 1D simulations presented in Section \ref{sec:backg}. The electron beam had a Gaussian profile in all directions, while the laser pulse had a Gaussian profile in the transverse direction and a polynomial profile in the longitudinal direction (see Appendix A). The focal spot, length, and width are specified as the $e^{-2}$ values.
}
\begin{ruledtabular}
\begin{tabular}{lc}
\textrm{Laser Pulse Parameters}&
\textrm{Value}\\
\hline
$\lambda_1 (\mu \mathrm{m})$ & 1 \\
$\lambda_2 (\mu \mathrm{m})$ & 0.5 \\
$a_1$ & 150 \\
$a_2$ & 75 \\
$\lambda_1$ \& $\lambda_2$ FWHM duration (fs) & 23\\
$\lambda_1$ \& $\lambda_2$ focal spot ($\mu \mathrm{m}$) & 6\\
Total Energy (J) & 800\\
\hline
\textrm{Electron Beam Parameters}&
\textrm{Value}\\
\hline
Length $(\mu \mathrm{m})$ & 6\\
Width $(\mu \mathrm{m})$ & 6\\
Charge (pC) & 10\\
Energy (GeV)  & 5.1\\
\end{tabular}
\end{ruledtabular}
\end{table}

In 2D (or 3D), the relative drift of the electrons and positrons eventually results in their spatial separation, which can facilitate experimental detection and diagnosis. To illustrate the spatial separation and demonstrate phase control for realistic, focused laser pulses, the head-on collision of a two-color laser pulse and a relativistic electron beam was simulated using 2D QED-PIC (see Appendix A for details). The physical parameters are motivated by near-term experimental capabilities and are displayed in Table \ref{tab:table1}.\cite{ELI,Apollon,CoReLS,EP-OPAL} The simulated interaction was synchronized so that the peak intensity of the laser pulse and the peak density of the electron beam arrived at the focal plane ($z=0$) at the same time. Note that the number of photons generated in NCS, and hence the number of pairs, is linearly proportional to the number of beam electrons. Thus, the pair and current densities can be linearly scaled to obtain the result for higher or lower beam charges. 

\begin{figure}[!ht]
    \includegraphics[width=\columnwidth]{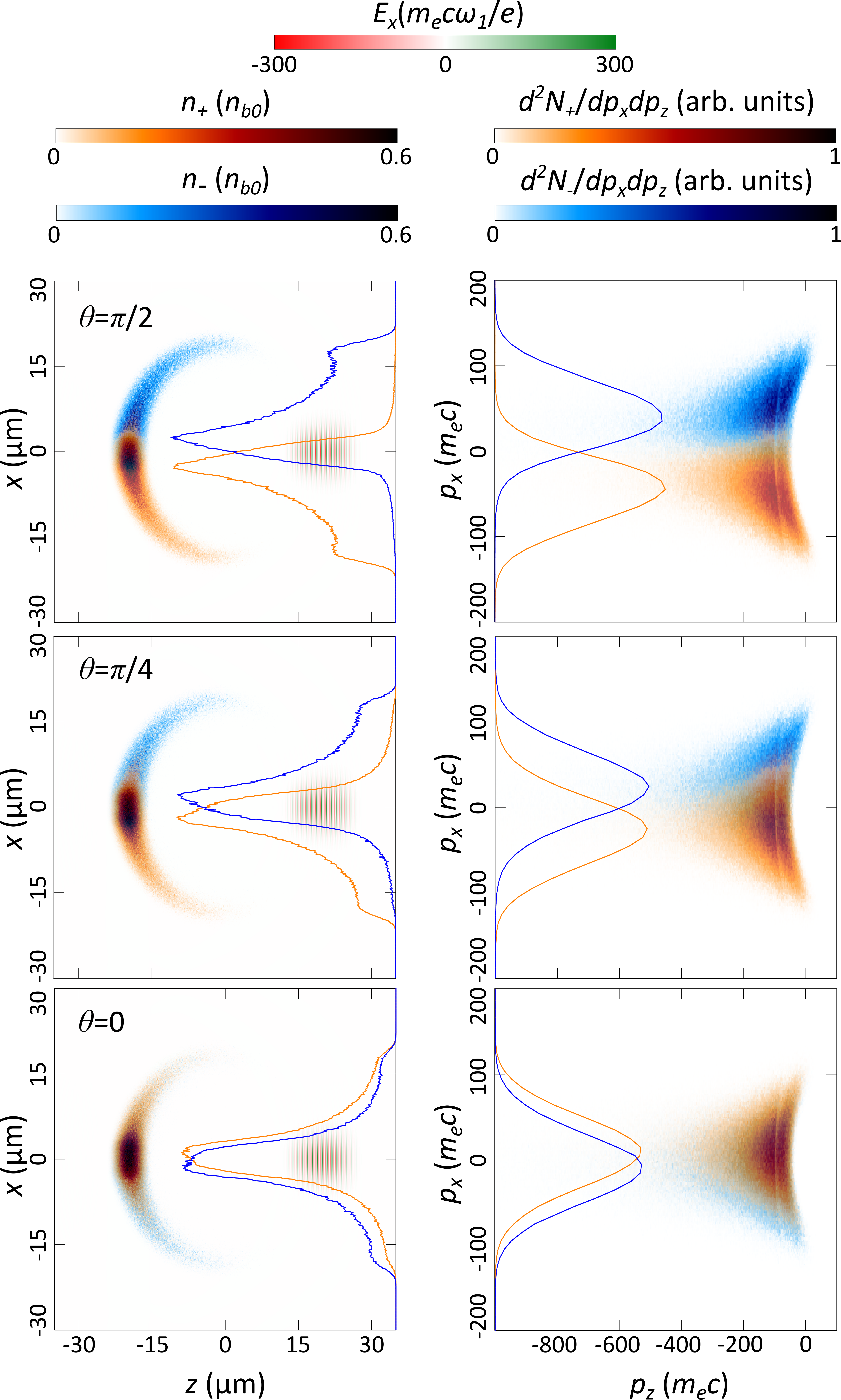}
    \caption{Charge densities (left) and momentum distributions (right) of electrons (blue) and positrons (orange) created by the nonlinear Breit-Wheeler process $\sim$65 fs after the collision of a two-color laser pulse and relativistic electron beam. A phase offset of $\theta = \pi/2$ (top) between the first and second harmonic maximizes the spatial separation and relative drift of the electrons and positrons. The separation and relative drift are smaller for $\theta = \pi/4$ (middle) and are minimal for $\theta = 0$ (bottom). The line outs show the density and momentum distributions integrated over $z$ and $p_z$, respectively.}
    \label{fig:the_main_image_of_this_paper}
\end{figure}

Figure ~\ref{fig:the_main_image_of_this_paper} displays the charge densities and momentum distributions of the created electrons and positrons $\sim$65 fs after the initial collision of the laser pulse and the electron beam. A phase offset of $\theta = \pi/2$ maximizes the relative drift (transverse momenta) and spatial separation of the electrons and positrons: The electrons travel predominantly in positive $x$ direction and the positrons in the negative $x$ direction. As the charges advance into the ``far-field,'' their relative drift will continue to increase their spatial separation. In accordance with the analytical model, the relative drift is smaller for $\theta = \pi/4$ and smaller still for $\theta = 0$. When $\theta = 0$, the electrons and positrons have a slight drift in the opposite direction (negative $x$ and positive $x$, respectively). This results from the difference in the Gouy phase of the first and second harmonic.

\begin{figure}[!ht]
    \includegraphics[width=\columnwidth]{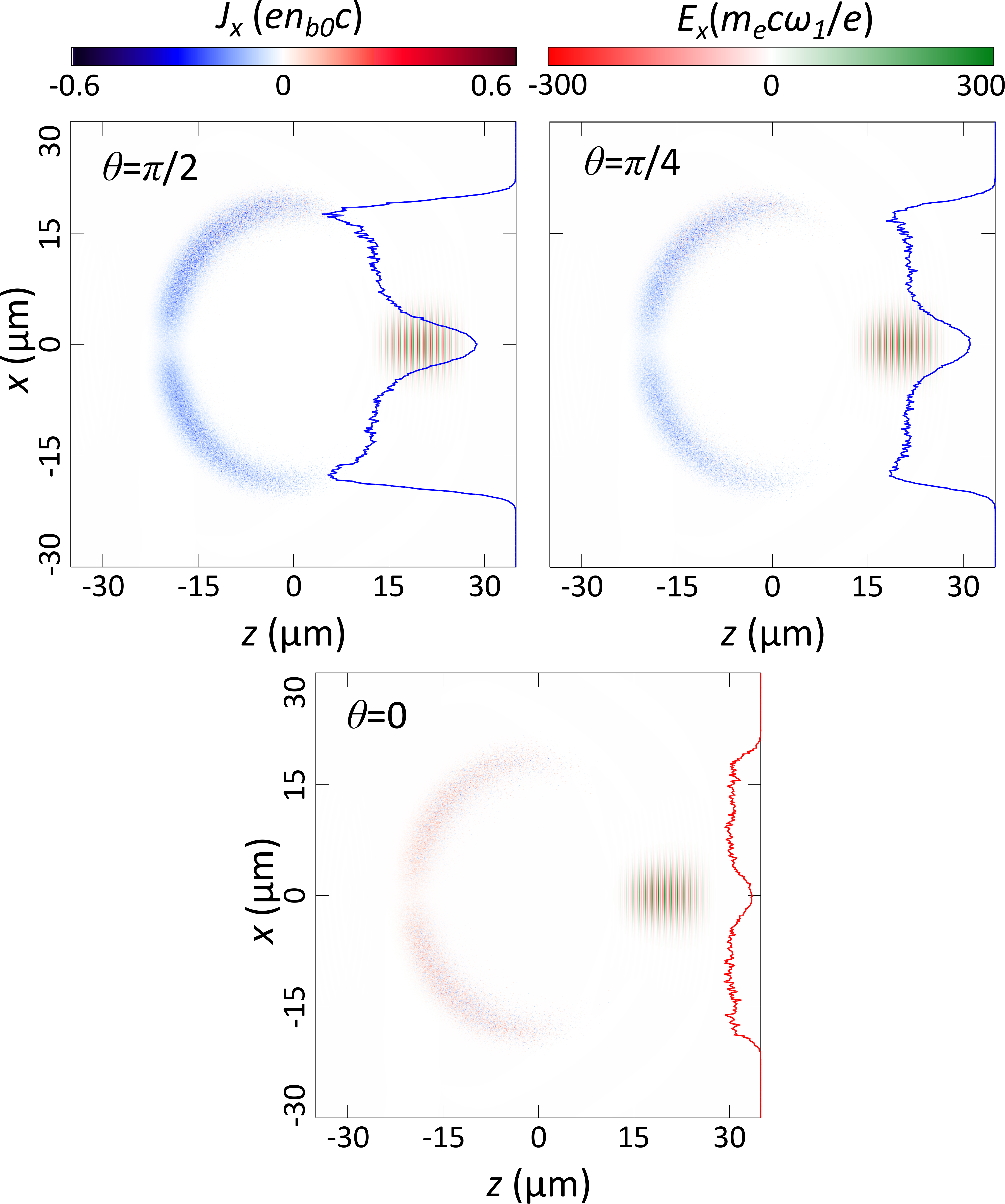}
    \caption{Transverse current densities for the cases displayed in Fig.~\ref{fig:the_main_image_of_this_paper}. The magnitude of the current density drops as the phase offset is decreased from $\theta = \pi/2$ to $\theta = 0$. The line outs show the integral of the current density over $z$. The large current density in the case of $\theta = \pi/2$ may provide a collective signature of nonlinear Breit-Wheeler pair creation.}
    \label{fig:currents}
\end{figure}

Figure~\ref{fig:currents} shows the corresponding transverse current densities for the cases displayed in Fig.~\ref{fig:the_main_image_of_this_paper}. When $\theta=\pi/2$, the positive transverse drift of the electrons and the negative transverse drift of the positrons produce two lobes of current located symmetrically about the propagation axis. Near the propagation axis, the current density is nullified by the spatial overlap and near-equal drift of the charges. When $\theta=\pi/4$, the situation is similar, but the charges are more overlapped in space, resulting in a much smaller current density. When $\theta=0$, the current density is barely distinguishable from noise. 

\section{Conclusions and Prospects}
\label{sec:conc}

A laser pulse composed of a fundamental and second harmonic provides control over the relative transverse motion of electrons and positrons created in the nonlinear Breit-Wheeler process. By adjusting the relative phase $\theta$ of the harmonics, the electrons and positrons can be made to drift in the opposite ($\theta = \pi/2$) or same direction ($\theta = 0$). In the case of $\theta = \pi/2$, the opposite drift spatially separates  the charges, which can facilitate experimental detection and diagnosis of the pairs. The opposite drift also produces a transverse current density that can drive low-frequency radiation. This radiation may offer an additional diagnostic signature of nonlinear Breit-Wheeler. Finally, the asymptotic angle of the pairs $\sim$$p_{xf}/p_{zf}$ depends on the vector potential at the time of creation a$_{x0}$, which could provide an indirect measurement of its value (Eqs. \ref{eq:conservations} and \ref{eq:drift}).

The underlying physics allowing for phase control was demonstrated using an analytical model and QED-PIC simulations with parameters motivated by near-term, high-power laser facilities. Second harmonic generation of a high-power ($>5$ PW) laser pulse has yet to be demonstrated and may be a technological challenge. Second (and third) harmonic generation of high-\textit{energy} laser pulses is now routine at long-pulse facilities like OMEGA and NIF. \cite{OMEGA,NIF} At a high-\textit{power} laser facility featuring two laser pulses, such as the proposed EP-OPAL laser, \cite{EP-OPAL} a 2nd harmonic conversion crystal could be placed before the final compression grating of one of the pulses. However, additional research is required to determine the feasibility of this approach. For example, the spectral acceptance of the frequency conversion crystal must accommodate the bandwidth of the short pulse, and high-efficiency compression gratings for $\sim$500 nm wavelengths would need to be fabricated.

This manuscript focused solely on the relative phase between two harmonics, but additional structuring of the optical waveform would allow for more exotic electron-positron motion. For instance, a two-color laser pulse with a time-dependent polarization could produce intertwined helical or twisted drifts. As another example, a two-color laser pulse with a higher-order transverse mode or orbital angular momentum could be used to spatially shape the drift motion and transverse current density. Finally, a two-color ultrashort flying focus \cite{Palastro2020,Ambat2023,Pigeon2023} would provide control over the velocity of the pair-creation front, enabling a quasiparticle source of superradiant emission. \cite{Malaca2023}

\begin{acknowledgments}
The authors would like to thank Dr. Bertrand Martinez and Dr. Christophe Dorrer for many insightful and useful discussions. The authors acknowledge the support of the Portuguese Science Foundation (FCT) Grant No. PTDC/FIS-PLA/3800/2021 as well as PRACE for awarding access to MareNostrum based in the Barcelona Supercomputing Centre and LUMI based in the data center of CSC in Kajaani, Finland. The work of JPP, DR, and KW is supported by the Office of Fusion Energy Sciences under Award Number DE-SC00215057, the Department of Energy National Nuclear Security Administration under Award Number DE-NA0003856, the University of Rochester, and the New York State Energy Research and Development Authority.
\end{acknowledgments}

\section*{Data Availability Statement}

The data that support the findings of this study are available from the corresponding author upon reasonable request. 

\section*{Conflict of Interest}
The authors have no conflicts to disclose.

\section*{Author Contributions}

BB conducted and analyzed the OSIRIS simulations. The analytical model was developed by BB, JPP, and MV. The manuscript was written by BB and JPP, with input from MV, KW, and DR. MV and JPP supervised the project. All authors discussed the results presented in the paper.

\appendix
\section{Simulation Details}
All simulations were run using the \textsc{osiris} framework\cite{OSIRIS} with the Monte Carlo module for QED processes. \cite{Gonoskov2015,Ridgers,ClassicalRR,Rates1,Rates2,Rates3}  The ``1D'' simulations modeled a plane wave laser pulse in a 2D domain with periodic boundary conditions in the transverse direction. The domain was $70~\mathrm{\mu m}~\times~10~\mathrm{\mu m}$ in the longitudinal ($z$) and transverse ($x$) directions, divided into 8600 $\times$ 320 cells. The asymptotic current density displayed in Fig.\ref{fig:comparisons} was obtained by averaging over the $x$ direction and summing over the $z$ direction. The 2D simulations modeled a focused laser pulse with open boundary conditions. In this case, the domain was $70~\mathrm{\mu m}~\times~60~\mathrm{\mu m}$, divided into 8640 $\times$ 1800 cells. For both the ``1D'' and 2D simulations, the electric fields of the harmonics were initialized with the temporal profile $f(t) = 10(-|t|+\tau_s)^3/\tau_s^3-15(-|t|+\tau_s)^4/\tau_s^4+6(-|t|+\tau_s)^5/\tau_s^5$, where $\tau_s=1.3\tau_{\mathrm{FWHM}} = 30~\mathrm{fs}$. A $27~\mathrm{attosecond}$ time step was used in all simulations, which ensured that the Courant-condition was satisfied and that in each time step the probabilities per particle for photon (NCS) and pair (nonlinear Breit-Wheeler) creation were much smaller than one. Thus, the probability that a single particle could be responsible for two creation events in a single time step was negligible. Four particles per cell were used for the initial electron beam in the 2D simulations of the focused pulse. 

\nocite{*}
\bibliography{bibliography}

\end{document}